\input{aipcheck} 
 
\documentclass[ 
    ,final            
  ] 
  {aipproc} 
 
\layoutstyle{6x9} 
 
\begin{document} 
 
\title{New Challenges to Hydrodynamics from Azimuthal Anisotropy at RHIC} 
 
\classification{25.75.-q,25.75.Ld} 
\keywords      {$v_{1}$, $v_{2}$, $v_{4}$, hydrodynamics} 
 
\author{A. H. Tang for the STAR Collaboration}{ 
  address={Physics Dept., Brookhaven National Laboratory, USA} 
} 
 
\begin{abstract} 
  This paper presents $v_{4}/v_{2}^2$ ratio as a function 
  of transverse momentum ($p_{t}$), pseudorapidity ($\eta$) and collision  
  centrality in Au+Au collisions at $\sqrt{s_{NN}} = 200$ GeV using the STAR  
  detector at the Relativistic Heavy Ion Collider (RHIC).  
  It is found that $v_{4}/v_{2}^2$ is larger than Hydrodynamic calculations,  
the centrality and transverse dependence of this ratio can not be fully described  
by Hydrodynamics, and the pseudorapidity dependence is opposite to what one  
expects from Hydrodynamics. The $p_{t}$ dependence of $v_{1}$ is also presented.  
It is found that $v_{1}(p_{t})$ for $|\eta|<1.3$ changes sign, and two possible  
explanations of the sign change are discussed. 
 
\end{abstract} 
 
\maketitle 
 
 
\section{Introduction} 
 
Azimuthal anisotropy  
describes the distribution of particles with respect to the reaction plane, 
and it is usually characterized by the Fourier harmonics ($v_{n}$) of this 
distribution~\cite{Posk98}. As a sensitive probe of the bulk properties of  
the system created in ultrarelativistic nuclear collisions, it has gained  
increasing attention from both experimentalists and theorists~\cite{STARPRC200GeV}. 
The large elliptic flow ($v_{2}$) observed at RHIC has been 
viewed as a success of Hydrodynamics in heavy ion collisions~\cite{STARFirst2FlowPaper},  
and the agreement of $v_2$ between RHIC data and Hydrodynamic predictions is 
used as one piece of evidence for the discovery of a dense and perfect  
liquid~\cite{science}. 
The recently measured higher harmonic $v_4$ provides new details regarding the  
shape of the distribution of particles in momentum space~\cite{v1v4}, and its magnitude  
and even its sign is important in understanding the initial configuration 
and the system evolution~\cite{Kolb}. In a more recent work, it is suggested that 
$v_{4}/v_{2}^2$ can be directly used as a probe of ideal fluid behavior~\cite{Borghini}. 
For a perfect fluid, this ratio is 0.5, and deviations from ideal-fluid 
behavior may yield higher values due to the increased value of $v_{4}$. It is thus  
generally expected that this ratio should become larger at high $p_{t}$ (>2 GeV/c) 
where Hydrodynamics breaks down. For the same reason, one also expects it to be  
larger at forward rapidity as well as in peripheral collisions. It is  
interesting to compare those expectations with data. 
As a successful model, Hydrodynamics is expected to describe not only even harmonics  
but also odd harmonics like directed flow ($v_{1}$), which characterizes the sideward  
bounce of particles, and carries very early information from the collision. In the  
past, most of  
the comparisons between Hydrodynamic calculation and data were focused on $v_2$ 
and $v_{4}$, and in this paper, we make the first attempt to compare $v_{1}$  
near midrapidity to Hydrodynamic predictions. 
\vspace{-0.5cm} 
\section{Results} 
 
\begin{figure}[ht]       
\resizebox{24pc}{!}{\includegraphics{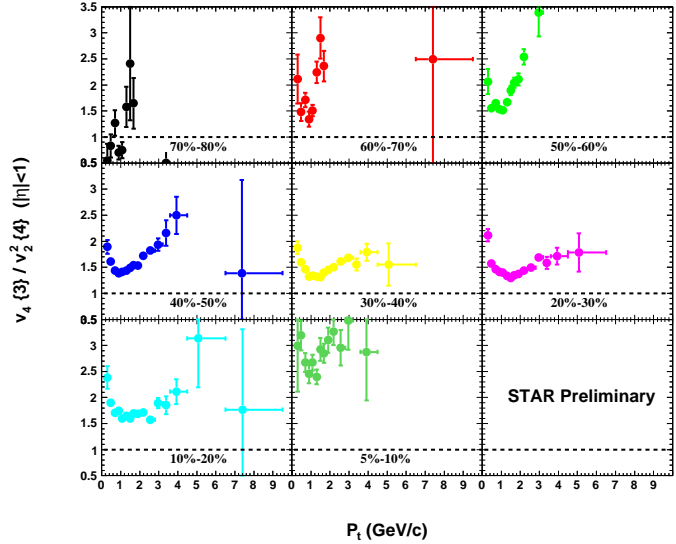}}      
\caption{\label{fig:ratioPt}  
$v_{4}\{3\}/v_{2}^2\{4\}$ as a function of $p_t$ for different centrality classes. 
} 
\end{figure}

Figure~\ref{fig:ratioPt} shows $v_{4}\{3\}/v_{2}^2\{4\}$, in Au+Au collisions at  
$\sqrt{s_{NN}} = 200$ GeV, as a function of $p_t$ for 8 centrality classes. The  
result for the top $5\%$ centrality is not shown because the computation of  
$v_{2}\{4\}$ fails for that centrality. In the plot, $v_{4}\{3\}$ and $v_{2}\{4\}$   
stand for $v_{4}$ measured with three-particle cumulants~\cite{3partCumu}, and  
$v_{2}$ measured with four-particle cumulants~\cite{4partCumu}, respectively. The  
non-flow effect, which is one of the largest sources of systematic uncertainty in  
azimuthal flow analyses, is suppressed significantly in both measurements.  
We found that the ratio decreases as $p_t$ increases until  
$p_{t} \sim 1 $ GeV/c where it reaches its minimum, and above $\sim$ 1 GeV/c  
in $p_{t}$, the ratio begins to increase. STAR reported 1.2 for the ratio  
$v_{4}/v_{2}^2$ in ref~\cite{v1v4}, and in that publication, the $v_{2}$  
used in the ratio was measured by the standard event plane method ($v_{2}\{EP_{2}\}$).  
The ratio is higher than 1.2 if one uses $v_{2}\{4\}$ instead of $v_{2}\{EP_{2}\}$,  
as shown in Fig.~\ref{fig:ratioPt}.  In the centrality and $p_{t}$ range we have  
studied, even the minimum of this ratio, which occurs at $p_{t} \sim 1 $ GeV/c in  
mid-central collisions, is far beyond the expected value of 0.5 from ideal  
Hydrodynamics. The rise of the ratio for $p_{t}$ beyond $\sim$ 1 GeV/c can be  
understood as the consequence of the breakdown of Hydrodynamics at high $p_{t}$,  
however so far there is no accurate prediction for the fall-and-rise feature of  
this ratio as a function of $p_{t}$.  
 
\begin{figure}[ht]       
\resizebox{24pc}{!}{\includegraphics{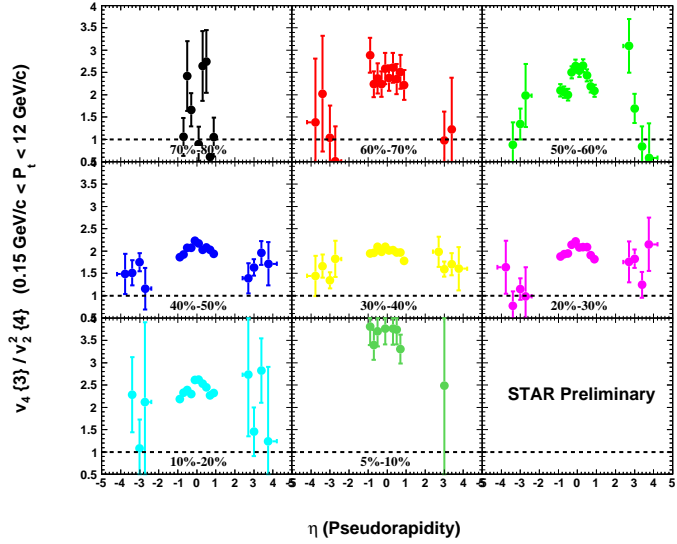}}      
\caption{\label{fig:ratioEta}  
$v_{4}\{3\}/v_{2}^2\{4\}$ as a function of $\eta$ for different centrality classes. 
} 
\end{figure}

The pseudorapidity dependence of this ratio is shown in Fig.~\ref{fig:ratioEta}. 
Because Hydrodynamics works best at mid-rapidity and breaks down at forward rapidity,  
this ratio is expected to be a minimum at mid-rapidity. For a similar reason, one  
would expect this ratio to decrease with centrality.  However, the data show the  
opposite trend --- the ratio is peaked at mid-rapidity, and the centrality dependence  
is more complicated, starting from peripheral events, it first decreases with centrality  
and reaches a minimum for mid-central events, then rises for central events. Note that  
the expectations mentioned above are based on pure Hydrodynamics, and there are other  
effects like viscosity that have not been taken into account. Other non-Hydrodynamic effects like the hadronic cascade at a later stage may also contribute to this ratio. In order to have a  
comprehensive understanding of this ratio, it is desirable to study the influence from these other factors.  
 
\begin{figure}[ht]       
\resizebox{35pc}{!}{\includegraphics{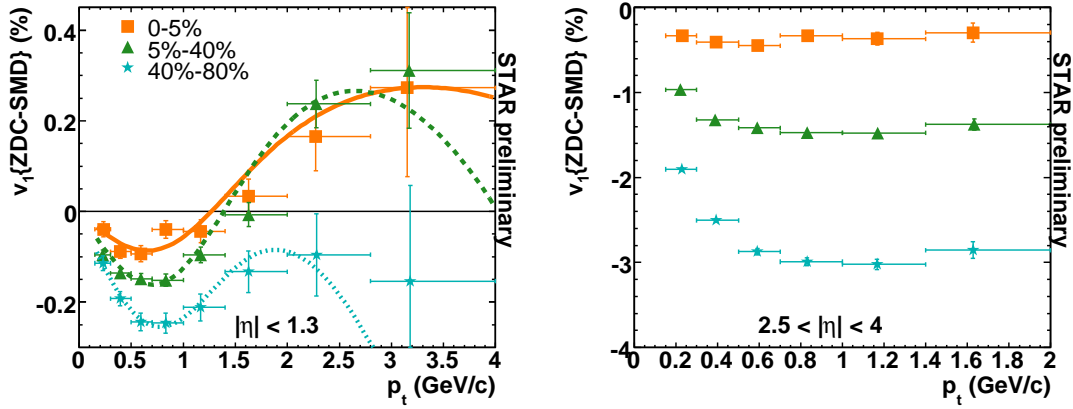}}      
\caption{\label{fig:v1pt}  
The left panel shows $v_{1}$ \{ZDC-SMD\} versus $p_{t}$ measured in the main TPC  
(|$\eta$|<1.3), for centrality 10\%-70\% in Au+Au collisions at $\sqrt{s_{NN}} =  
200$ GeV. The event plane was reconstructed using Shower Maximum  
Detectors incorporated in the Zero Degree Calorimeters (ZDC-SMD). The right panel  
shows $v_1$ \{ZDC-SMD\} versus $p_{t}$ measured in the forward TPCs (2.5<|$\eta$|<4.0),  
for different centralities in Au+Au collisions at $\sqrt{s_{NN}} = 200$ GeV. The sign  
convention is such that the directed flow of spectator neutrons at positive beam  
rapidity is positive.  The data points are from~\cite{Gang}. 
} 
\end{figure}

It was shown in a recent work~\cite{Heinz} that in a Hydrodynamic framework, the  
initial ``tilting'' in the transverse overlap region can make particles shift sidewards  
in the direction of the impact parameter, and give rise to a non-zero directed flow  
$v_{1}$. Because the colliding matter receives no overall transverse kick in this model,  
the $p_{t}$-integrated directed flow is zero. That means that if $v_{1}(p_{t})$ is  
non-zero at low $p_{t}$, it has to change its sign at a higher $p_{t}$. The  
$v_{1}(p_{t})$ in Au+Au collisions at $\sqrt{s_{NN}} = 200$ GeV is shown in  
Fig.~\ref{fig:v1pt}. 
We found that $v_{1}(p_{t})$ decreases with centrality. In central collisions, the  
$v_{1}(p_{t})$ near the mid-rapidity region ($|\eta|<1.3$) changes sign around  
$p_{t}=1.5$ GeV/c. The sign change and the $p_{t}$ where it happens is consistent  
with the prediction in~\cite{Heinz}. However, if protons and pions are flowing in  
the direction opposite to each other, a sign change can also be expected simply due  
to the enhancement of baryon to meson production at intermediate $p_{t}$. Taking the  
relative yield of pions and protons, we can fit the charged particle $v_{1}$, as  
shown by curves in the left panel of Fig.~\ref{fig:v1pt}, with good $\chi^2 / $ndf  
($0.2 - 0.5$). Currently we can not distinguish the two scenarios. The right panel  
is the same measurement obtained in the forward region ($2.5<|\eta|<4$). No crossing  
zero of $v_{1}(p_{t})$ is observed in that pseudorapidity range. 
 
\vspace{-0.2cm} 
\section{Summary} 
\vspace{-0.2cm} 
 
We have presented the ratio $v_{4}\{3\}/v_{2}\{4\}^2$ as a function of $p_t$ and  
$\eta$ for different collision centralities. We found that this ratio increases at  
high $p_{t}$ which is in line with Hydrodynamics, but there is no accurate prediction  
for the fall-and-rise feature of this ratio as a function of $p_{t}$. The ratio is  
found to be peaked at mid-rapidity, contrary to Hydrodynamics. It is also found to be  
a minimum for mid-central events, as opposed to decreasing monotonically with  
centrality, as expected from Hydrodynamic arguments. We found that $v_{1}(p_{t})$ for  
$|\eta|<1.3$ changes sign, consistent with a tilted initial overlap region followed  
by a Hydrodynamic evolution. The sign change could also be explained by enhanced  
baryon production at intermediate $p_t$, with pions and protons flowing opposite to  
each other. How to explain our findings is a new challenge to Hydrodynamic models. 
 
\vspace{-0.2cm} 
\begin{theacknowledgments} 
\vspace{-0.2cm} 
The author thanks Y. Bai, R. Snellings and G. Wang for their contributions. 
\end{theacknowledgments} 
\vspace{-0.2cm} 
 
\bibliographystyle{aipproc}   
 
 

\end{document}